# ENERGY SECURITY: KEY CONCEPTS, COMPONENTS, AND CHANGE OF THE PARADIGM

**Julia Edigareva,**
*Associate Professor, Russian University of Transport,*
*Moscow, Russia*

**Tatiana Khimich,**
*Associate Professor, Russian University of Transport,*
*Moscow, Russia*

**Oleg Antonov,**
*Associate Professor, Russian University of Transport,*
*Moscow, Russia*

**Jesus Gonzalez,**
*Lecturer, Complutense University of Madrid,*
*Madrid, Spain*

**Abstract**

The authors of the study conduct a legal analysis of the concept of energy security. Energy is vital for sustainable development, and sustainability is not only at the heart of development, but also of economic, environmental, social and military policies. To ensure the sustainability of the policies 'security' appears as a mandatory objective to achieve. The article critically assesses the change in the energy paradigm.

## 1. Introduction

The concept of the 'energy security' was first introduced by industrialized market economies in 1947, when the United States adopted a document regulating the actions of the state in the field of national security. Although it has been investigated and developed by a wide range of scholars and specialists, still there is no universal understanding of what is meant by this notion.

## 2. Main part

In international relations, the term 'energy security' acquired significance and became widely used after the global oil crisis of 1973-1974, when the world faced a sudden and noticeable increase in prices for fuel and energy resources caused by the following political events [1]:

- collective embargo imposed by the Arab OPEC countries on the export of oil to the USA;
- support of Israel in the October war by a number of European countries.

Being faced with the first largest energy crisis, industrial countries acutely felt their vulnerability and consequently energy security became one of the top priorities, considered a part of the country's national security.

As mentioned above, currently, despite the abundance of modern research and publications on energy topics, a single generally accepted definition is still not formed. In point of fact, the meaning of energy security differentiates from country's dependence to their energy imports. Accordingly, countries which are highly dependent on imported oil and gas adheres energy security to supply whereas, countries which export oil and gas adheres energy security to demand. This is due to the fact that different groups of countries, namely exporting countries, importing countries and transit countries interpret this concept in their own way, based on their national and economic interests and priorities.

Therefore, in order to give a clear definition of the concept of "energy security", it is imperative to consider it from the perspective of all three parties. But, nevertheless, in modern scientific literature there are many concepts and interpretations of international energy security, and representatives of different schools or countries offer their own definitions of this notion.

The International Energy Agency (IEA), which is an international body that unites 29 industrialized countries under the Organization for Economic Development and Cooperation, defines energy security as 'the uninterrupted availability of energy sources at an affordable price' [2]. The limitations of this approach are directly implied by the fact that in this context, it considers energy security from the point of view of importing countries that reflects their national priorities in the face of growing dependence on imports of oil and natural gas.

Nevertheless, the IEA admits that the concept of 'energy security' can be divided into two major groups: long-term energy security and short-term energy security. The first one embraces 'timely investments to supply energy in line with economic developments and environmental needs', while the latter deals with 'the ability of the energy system to react promptly to sudden changes in the supply-demand balance' [2].

Nowadays one of the major energy importing country is China that is primarily preoccupied with the security of the energy supply. Beijing defines 'energy security' as 'reliable uninterrupted supply of energy, which is necessary for economic development of the country' [1]. In other words, energy security is seen as a need for uninterrupted supply of energy resources in order to maintain stable economic growth.

Meanwhile, the exporting countries, including Russia, do not share this position. They are primarily interested in diversifying the markets of energy exports, preserving their national sovereignty and of course guaranteeing sustainable demand for the energy they export. Because of diverging interests in the energy market, exporting countries interpret 'energy security' in another way.

The OPEC group (a permanent intergovernmental organization of 15 oil-exporting developing nations that coordinates and unifies the petroleum policies of its Member Countries [4]) defines energy security as 'reliable and environmentally friendly energy supplies at prices reflecting fundamental principles of market economy'. This approach is also limited, as the IEA one, because it reflects only the interests of exporting countries that are focused on determining a decent price for energy resources.

The interests of transit countries consist mainly in maximizing transit rent of energy resources. This is the approach Turkey follows, - the country is a natural energy corridor between the Middle East, the Caspian basin and Europe and contributes significantly to the attempts of Europe to diversify its energy suppliers for natural gas.

In the meantime, the concept of energy security framed as an access to fossil fuels created the so-called 'energy nationalism' [3] that created a reality in which the behavior and decisions made on energy markets and the delivery of resources ultimately depended not on the economic market factors, but on producers, which resulted in the transformation of the energy market into the subject of inter-state relations. Oil and natural gas were used as geopolitical weapons, and geopolitics and geo-economics became an important part of world politics and foreign policy ofthe main players on the energy market.

Energy security undoubtedly depends on the national and international background. In the context of state-controlled markets, the main guardians of energy security are the governments. On the contrary, given the energy markets are liberalized and the main actors are private companies, the security of supply consists of an efficient risk management strategy by governments, companies and consumers. Therefore, it is important not to fall into purely ideological approaches.

Since Adam Smith, the state is authorized to intervene in the economy in order to prove security, including its energy component, to its citizens. Baring that in mind,it should be noted that such intervention should be done as far as possible in a concerted manner along with the companies and consumers, following the principle of subsidiarity.

At the same time, energy security is associated with the evolution of the energy markets, the geopolitical situation and long-term international scenarios. There are several scenarios drawn for the long term that extrapolate to some extent conceptual differences, for example, the 'Markets and Institutions' scenario and its alternative called 'Empires and Regions' [3], and the three scenarios drawn by Shell Global Scenarios [5]: 'Low Trust Globalization', 'Open Doors' and 'Flags'. The 'Low Trust Globalization' is based on the trade-off between efficiency and security, and is characterized by limited international integration, intrusive state intervention and institutional discontinuities. The 'Open doors' arises from the dilemma between efficiency and environmental sustainability, premium market incentives and the participation of the civil society. It highlights the urgent necessity of transnational integration and harmonization, and mutual recognition of standards. The 'Flags' responds to the mercantilist logic and involves regulatory fragmentation, nationalism and the conflict between the values of the different regions.

To some extent, the differentiation of scenarios roots in the neoliberal and neorealist paradigms to international energy issues. It is an old debate between two alternative visions of the world: a world in which the market disturbances are resolved by cooperation, or a fragmented world conceived as a billiard table where conflicts are resolved through the exercise of the political, economic and military power. Under the neoliberal paradigm, that is so praised by the EU, energy security is achieved through the development of markets and the management of conflicts at the multilateral level through supranational institutions. The neorealist paradigm of energy security, instead, implies the development of bilateral relations and the subordination of markets to foreign policy.

The potential risks and threats related to energy security arise mainly from two circumstances:

- the projected upcoming peak in the production of hydrocarbon resources, which is vital for the modern economy, and
- the security of their supplies.

However, recently the energy sector has started to develop a few key trends that were caused by new, very strong factors: the global financial and economic crisis and the shale revolution in the production of oil and gas. Today, energy security policies require a paradigm shift and a new model of factors and conditions for its implementation.

The first factor that radically changed the context of energy policy was the global economic crisis. Since 2008, experts have determined it as the financial crisis, the economic crisis, the crisis of democracy and governance, the crisis of the culture of public consumption and material culture in general, and the environmental crisis that will eventually lead to global natural disasters. It would be reasonable to say that the world experienced a multidimensional global crisis, or the first systemic crisis of the global era. In the energy sector, this crisis coincided with the start of the gradual transition from 'industrial' and 'hydrocarbon' to 'neo-industrial' and 'smart' energy, which includes the following aspects: smart grids, energy efficiency (in a broad sense), renewable energy, new principles of energy systems and focus shift from producers to consumers [3].

The second factor that dramatically changed the energy markets was the quiet shale revolution in oil and gas production.

The shale revolution, which has become a reality in the United States and Canada, as estimated by the experts, will have serious consequences for the global energy market. The unconventionally produced natural gas fundamentally changed the world market. The most serious consequence of the shale gas revolution is a shift in the focus from producers to consumers.

In the context of the old paradigm, energy security is directly related to energy independence. The idea was that if a country is self-sufficient with respect to energy resources, and has an efficient (energy-saving) economy, then this will be accompanied with lower prices for energy carriers. The reality of oil prices in the United States after the shale boom has shown that this is utopia.

### 3. Conclusion

Meanwhile, achieving self-sufficiency in the energy sector is almost impossible. Even such countries as Russia, Saudi Arabia, Venezuela, Brazil and Canada, which are rich in hydrocarbons, import a part of the energy in the form of refined petroleum products due to insufficient refining opportunities. This dependence could theoretically be eliminate with a little effort and investment in the construction of new refineries, but in practice it is not happening. Revolutionary changes of the energy security require a paradigm shift, whichshould be reflected in the energy security policies.

Currently we are on the verge of transition to a post-industrial, 'smart' energy system, which means 'smart' networks, alternative energy sources for transport, decentralization of energy, integration of energy into the technical sphere, accompanied by an increase in energy efficiency.

To conclude, it can be said that the definition of 'energy security' can be complete only if the interests of all the energy market participants are taken into account, and namely importing countries, exporting countries and transit countries. Interests and priorities of all three parties are different, which presents a difficulty in agreeing on a common concept. At the same time, with the paradigm shift all of the actors of the energy market are interested in ensuring the reduction of geopolitical and environmental risks and thus creating new opportunities.

**References**

[1] Correljé y Van der Linde (2006) Energy Supply Security and Geopolitics: A European Perspective. In *Energy Policy* nº 34.

[2] International Energy Agency [Electronic resource] URL: https://www.iea.org/weo/

[3] Kozlovsky V. V., Lutokhina E. A. (2014) World energy: scenario forecasts of development. - Moscow: IVC Ministry of Energy, 2014.

[4] OPEC official website [Electronic resource] URL: https://www.opec.org/

[5] The Shell Global Scenarios to 2025. The Future Business Environment: Trends, Trade-offs and Choices. [Electronic resource]. URL: https://rjohnwilliams.files.wordpress.com/2016/02/shell-global-scenarios2025summary2005.pdf